\documentclass{mn2e} \usepackage{epsfig,psfig}

\title[Analysis of the thin layer of Galactic warm ionized gas]
{Analysis of the thin layer of Galactic warm ionized gas in the
range $20^{\circ} < l < 30^{\circ}$, $-1.5^{\circ} < b <
+1.5^{\circ}$}
\author[R. Paladini, G. De Zotti, R.D. Davies, M. Giard]
{R. Paladini$^{1}$\thanks{E-mail: paladini@cesr.fr}, G. De
Zotti$^{2}$\thanks{E-mail: dezotti@pd.astro.it}, R. D.
Davies$^{3}$\thanks{E-mail:rdd@jb.man.ac.uk} M.
Giard$^{1}$\thanks{E-mail: giard@cesr.fr}\\ $^{1}$CESR,
9, Avenue du Colonel Roche, Boite postale 4346, F-31028 Toulouse, France\\
$^{2}$INAF--Osservatorio Astronomico di Padova, Vicolo
dell'Osservatorio 5, I-35122 Padova, Italy\\
$^{3}$University of Manchester, Jodrell Bank
Observatory, Macclesfield - Cheshire SK11 9DL, UK
}
\begin{document}

\pagerange{\pageref{firstpage}--\pageref{lastpage}} \pubyear{2003}

\maketitle

\label{firstpage}

\def\lsim{\,\lower2truept\hbox{${<\atop\hbox{\raise4truept\hbox{$\sim$}}}$}\,}
\def\gsim{\,\lower2truept\hbox{${> \atop\hbox{\raise4truept\hbox{$\sim$}}}$}\,}

\begin{abstract}

\noindent We present an analysis of the thin layer of Galactic
warm ionized gas at an angular resolution $\sim 10'$. This is
carried out using radio continuum data at 1.4 GHz, 2.7 GHz and 5
GHz in the coordinate region $20{\degr} < l < 30{\degr}$,
$-1.5{\degr} < b < +1.5{\degr}$. For this purpose, we evaluate the
zero level of the 2.7 and 5 GHz surveys using auxiliary data at
2.3 GHz and 408 MHz. The derived zero level corrections are
$T_{\rm zero}(2.7\,\hbox{GHz})=0.15 \pm 0.06\,$K and T$_{\rm
zero}(5\,\hbox{GHz})=0.1 \pm 0.05\,$K. We separate the thermal
(free-free) and non-thermal (synchrotron) component by means of a
spectral analysis performed adopting an antenna temperature
spectral index $-2.1$ for the free-free emission, a realistic
spatial distribution of indices for the synchrotron radiation and
by fitting, pixel-by-pixel, the Galactic spectral index. We find 
that at 5 GHz, for $|b| = 0\degr$, the fraction of thermal
emission reaches a maximum value of 82$\%$, while at 1.4 GHz, the
corresponding value is 68$\%$. In addition, for the thermal
emission, the analysis indicates a dominant contribution of the
diffuse component relative to the source component associated with discrete HII regions.

\end{abstract}


\section{Introduction}

The spatial distribution of the Galactic Warm Ionized Medium (WIM)
is characterized by a thin and a thick layer. The {\it{thin
layer}} consists of ionized gas located on the 
Galactic plane and composed of discrete HII regions and
diffuse gas. The {\it{thick layer}} consists instead of the
ionized gas which is situated well above the plane and presents
only a diffuse-gas component. Current models of the WIM (Miller
$\&$ Cox 1993; Domg\"{o}rgen $\&$ Mathis 1994)  tend to favour a
scenario in which the presence of ionized gas in the thick layer
is due to ultraviolet radiation leaking out of regions of star
formation located in the thin layer. In the light of these models,
knowledge of the physical properties of the thin layer is crucial
for understanding the mechanisms regulating the global
distribution of the warm ionized gas. However, despite its
relevance, systematic investigations of the thin layer have not
been carried out to date. Complementary information on the WIM
comes from pulsar dispersion measurements (DMs) which are linearly
weighted by the electron density, $n_{e}$, rather than weighted by
$n_{e}^2$ as in the case of the $H\alpha$ and of the radio
free-free emission discussed here. The DM data imply an inner
Galactic electron layer of scale height $\sim 150\,$pc (Taylor
$\&$ Cordes 1993); these data are spatially undersampled however
and do not have the angular resolution of the data used in the
present study.

This paper investigates the thin layer by combining the
information on the diffuse component derived from radio continuum
surveys with the information on the Galactic HII region
distribution as given in Paladini, Davies $\&$ De Zotti (2003). In
particular, we estimate the contribution of the discrete, HII
component with respect to the diffuse component. We emphasize that our 
analysis is also of interest for foreground studies in the 
context of CMB observations as well as for investigations of 
the ISM. The paper is organized as follows. In Sect.~2 we present the data
base used for the analysis. In Sect.~3 we discuss the importance
of an accurate determination of the zero levels for the surveys
involved and we describe the methods applied to our case. In
Sect.~4 we illustrate the results of the component separation
technique in deriving the latitude distribution of free-free and
synchrotron emission. In Sect.~5 we estimate the contribution of
discrete HII regions to the total free-free emissions.

\section[]{Choice of the data base}

For the analysis of the diffuse ionized gas adjacent to the
Galactic plane, we make use of radio continuum data. As well
known, an ionized gas in the physical conditions of the
interstellar medium ($T_{e} \sim 10^{4}$ K, $n_{e} \sim$ 0.1
cm$^{-3}$) emits a continuum spectrum at radio frequencies due to
Coulomb interaction of free electrons with ions (thermal
bremsstrahlung or {\em free-free radiation}). The use of radio data offers
various advantages with respect to other tracers: this continuum
radiation is not subject to extinction from interstellar dust
grains (as the H$\alpha$ emission) and it does not suffer from
undersampling (as in the case of pulsar dispersion measures). An
important drawback, however, for radio continuum data results from
the fact that the observed emission is the superposition of the
free-free emission and of the synchrotron radiation produced by
relativistic electrons accelerated in the Galactic magnetic field.
Therefore, the use of such data for investigating the ionized gas
requires a decomposition of the observed signal into its
constituents. This can be achieved by exploiting the spectral
dependence of each component when data are available at least at
two frequencies.

The first step is the selection of a region of the Galaxy in which
to carry out the analysis. This is followed by the choice of the
data sets covering the selected region. For the selection of the
Galactic region we follow these guidelines: we look for an area
surveyed at several frequencies, close to the plane and in a
longitude range where we expect the emission from the ionized gas
to be particularly bright. At the same time, we try to avoid
regions in which the source of emission is not fully understood
such as areas located in the proximity of the Galactic centre.
According to these criteria, we select the longitude range
$20{\degr} < l < 30{\degr}$ which includes the Sagittarius-Carina
and the Scutum-Crux arms; the latter is tangent at $l \sim
30{\degr}$. The Galactic plane emission is well-defined in this
region.

The three surveys with accurate calibrations and well-determined
baselines are the 1.4 GHz survey by Reich et al. (1990a), the 2.7
GHz survey by Reich et al. (1990b) and the 5 GHz survey by Haynes
et al. (1978). They cover a sufficient frequency range for an
adequate separation of synchrotron and free-free emission
components. It is also important to stress that the angular 
resolution of these surveys enables us to perform an analysis 
of the warm ionized gas in the Galactic plane on a 10$'$ scale. 
It is
for this reason that we have not used the 408 MHz data of Haslam
et al. (1982) at 51$'$ resolution. Our study, as a consequence,
covers the latitude range $-1.5{\degr} < b <1.5{\degr}$ defined by
the 5 GHz survey.

All the maps used for this work have been downloaded from the
MPIFR web site: http://www.mpifr-bonn.mpg.de/survey.html. Details
about the surveys are given below and summarized in Table~1.

\begin{table}
\caption[]{Summary of the surveys }
\begin{center}
\begin{tabular}{lccc}
\hline
\hline
{\it Survey} & $\nu$  & HPBW & Sky coverage \\
\hline
Effelsberg 100-m &  1.4 GHz &  9.4$'$ & $-3{\degr} < l < 240{\degr}$\\
                &         &        &   $-4{\degr} < b < 4{\degr}$\\
Effelsberg 100-m &  2.7  GHz &  4.3$'$ & $-2{\degr} < l < 240{\degr}$\\
                &         &        &     $-5{\degr}< b < 5{\degr}$\\
Parkes 64-m          &  5 GHz  & 4.1$'$ & $ -170{\degr} \le l \le 40{\degr}$     \\
                 &         &        &$-1.5{\degr} \le b \le 1.5{\degr}$ \\
\hline
\hline
\end{tabular}
\end{center}
\end{table}

\begin{figure*}
{\epsfig{figure=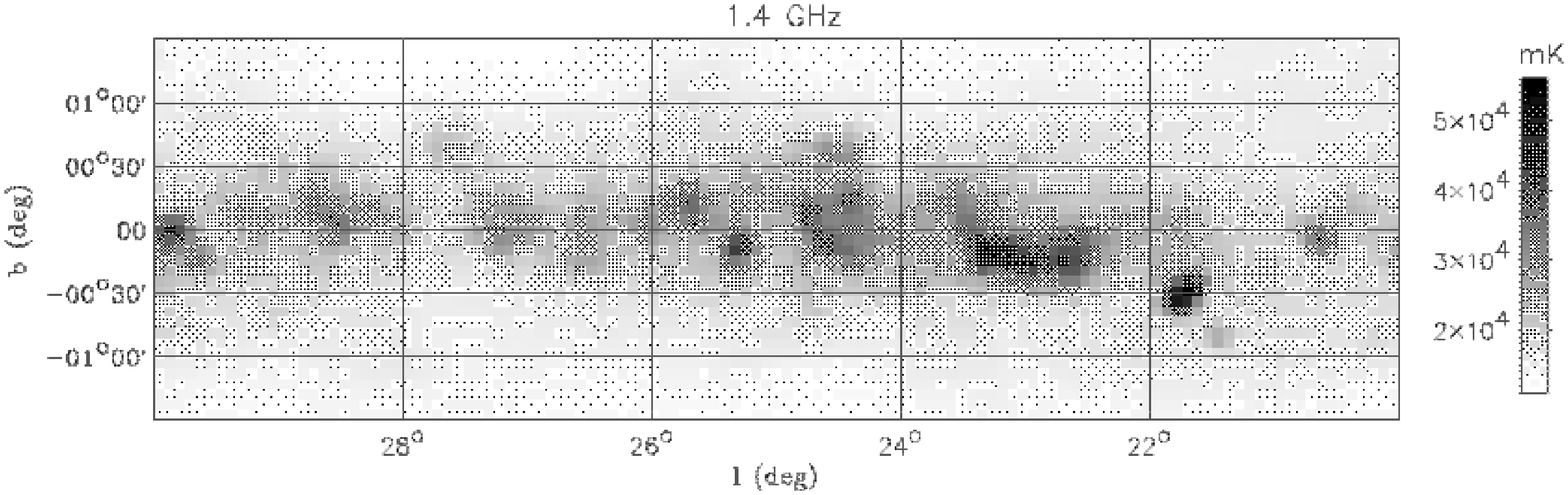, height=5cm, width=16.8cm, angle=0}\\
\vspace*{0.3truecm}
\epsfig{figure=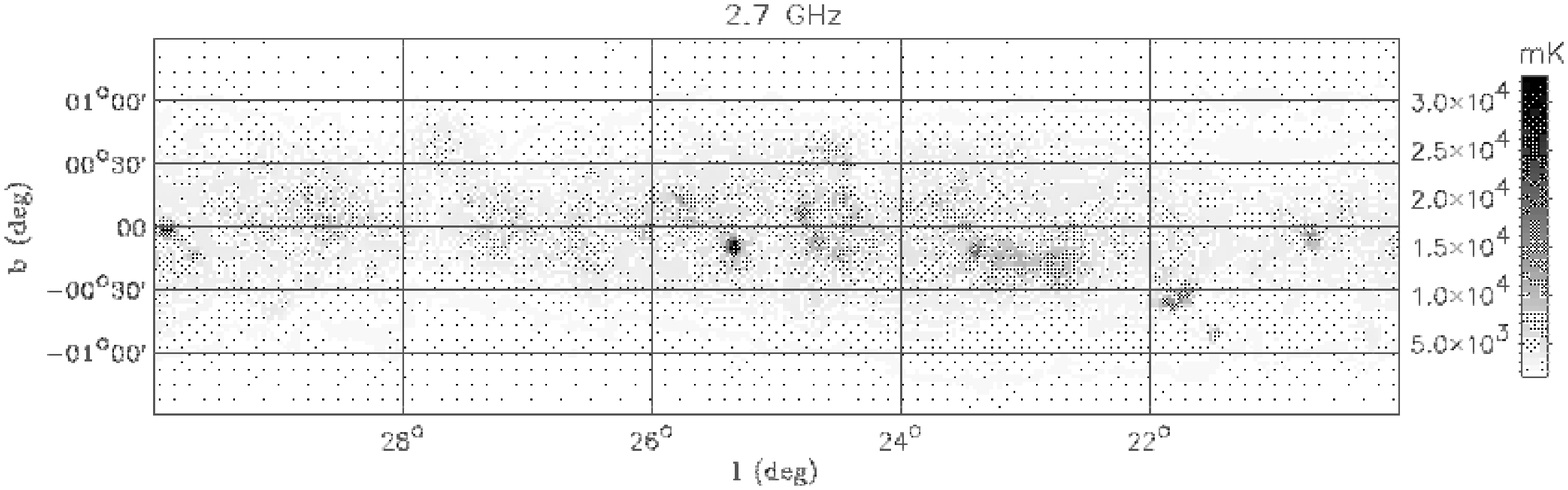, height=5cm, width=16.8cm, angle=0}\\
\vspace*{0.3truecm} 
\epsfig{figure=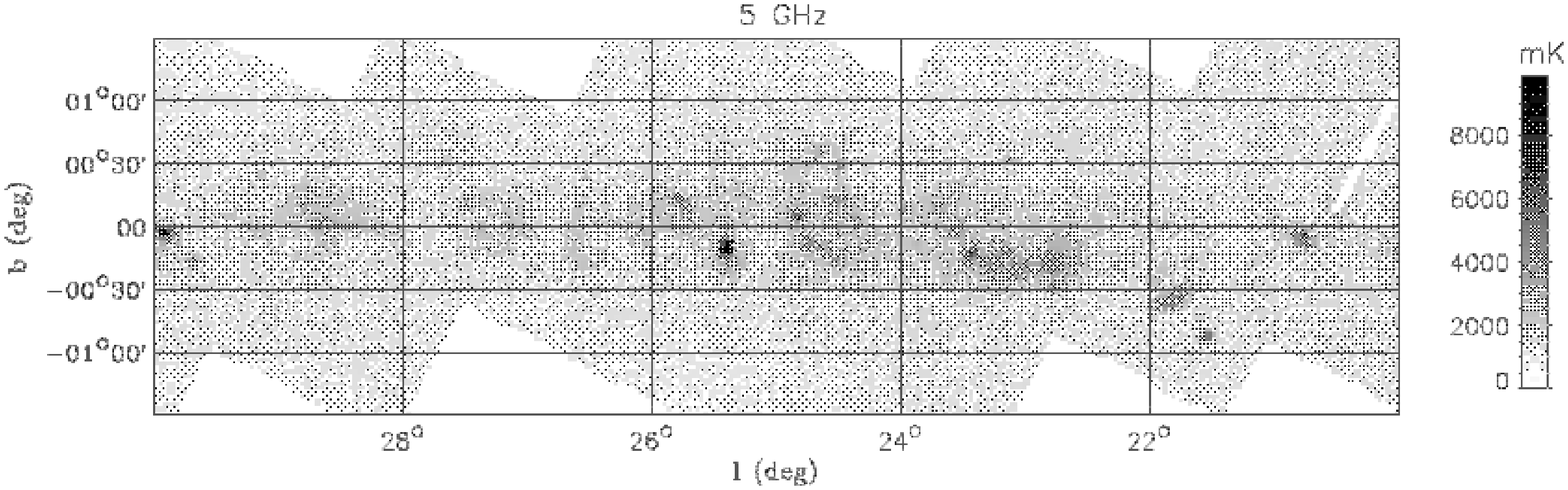, height=5cm, width=16.5cm, angle=0}} 
\caption{The selected
Galactic region $20{\degr} < l < 30{\degr}$ and $-1.5{\degr} < b <
+1.5{\degr}$ in the 1.4 GHz survey (top panel), 2.7 GHz survey
(central panel) and 5 GHz survey (bottom panel). The background
due to the CMB and extragalactic sources has been removed.}
\end{figure*}

\subsection{The 1.4 GHz survey}

The 1.4 GHz data are derived from a continuum survey of the 
Galactic plane carried out with the Effelsberg 100-m telescope 
at an angular resolution of 9.35$'$ in the coordinate range 
$-3{\degr}< l < 240{\degr}$, $-4{\degr} < b < 4{\degr}$. The
baseline level of the total intensity map has been determined from
the large area Stockert 1.4 GHz survey (Reich 1982; Reich $\&$
Reich 1986) made with a resolution of 35.9$'$. The absolute level
of emission is derived by comparison with the Howell $\&$
Shakeshaft (1967) measurements and is believed to be accurate to
0.5 K. The temperature calibration based on standard radio 
sources is accurate to $\pm 5\%$. The Reich et al. (1990a) survey,
corrected for the CMB and unresolved extragalactic sources (see
Section 3) is adopted as the reference survey for the present
analysis. It is plotted in Fig.~1.

\subsection{The 2.7 GHz survey}

The 2.7 GHz data have been obtained with the Effelsberg 100-m
telescope. The survey is characterized by an angular resolution of
4.3$'$ and covers the coordinate range $-2{\degr} < l <
240{\degr}$, $-5{\degr}< b < 5{\degr}$. The latitude range
$-1.5{\degr}< b < 1.5{\degr}$ was scanned in Galactic latitude
with a scan separation of 2$'$. The latitude range $|b| =
1.5{\degr}$ to 4${\degr}$ was scanned at the same scan separation
but in both Galactic coordinates. For $-2{\degr}< l < 76{\degr}$
the large-scale structures have been added by using low-resolution
data ($\sim 19'$) at 2.7 GHz taken with the Stockert telescope
(Reif et al. 1987). Zero level fluctuations are reported of order
of 100 mK. The Effelsberg data and the low-resolution data have
been combined into a final map. The temperature calibration
accuracy is believed to be $\pm 5\%$. The 2.7 GHz data corrected
for the CMB  and extragalactic sources are plotted in Fig.~1.

\subsection[]{The 5 GHz survey}

The 5 GHz survey was made using the Parkes 64-m telescope and has
an angular resolution of $4.1'$. It covers $-170{\degr} \le l \le
40{\degr}$ and $-1.5{\degr} \le b \le +1.5{\degr}$. The primary
scan direction was chosen to be either right ascension or
declination, whichever was nearer to being orthogonal to the
Galactic plane. In particular, the longitude range $20{\degr} \le
l \le 30{\degr}$ was scanned in right ascension, with primary 
scans separated by 2$^{\prime}$. The region of the sky to be
surveyed was divided into blocks, each block being $\sim$ 16
square degrees in area. Baseline levels within each block were set
by reference to two tie-down scans to points well off the Galactic
plane so as to determine a consistent base level. The baseline
uncertainty was estimated to be 0.2 K and the temperature accuracy
was $\pm 10\%$.

\vspace*{0.5truecm} \noindent A few steps of data processing have
been taken prior to the realization of the component separation. 
The 5 GHz map is characterized by a significant number of negative 
pixels -- typically at the border of the map -- which correspond 
to unobserved regions and result from the adopted scanning 
strategy. No interpolation over these bad pixels has been
performed since these regions are far too extended to allow us to
fill the gaps in a realistic, reliable way. We have therefore
carried out our analysis considering, for each frequency, only the
pixels with a positive, detected signal. All the maps have been
pixelized at the level of the 5 GHz map, i.e. to $2{'}$ pixel and
the 2.7 and 5 GHz maps have been convolved to 9.4$'$, the angular
resolution of the 1.4 GHz map. Due to the missing pixel problem,
the convolution introduces some level of error in the 5 GHz map.
However, this error concerns only a very limited number of pixels
(namely, the ones at the scan borders) and we expect it not to be
dominant in the global error budget.

\section[]{Zero level determination}

As mentioned in the introduction, it is possible to perform the
separation of the free-free emission from the synchrotron emission
by exploiting the information on their spectral behaviour.
However, this method is subject to large errors if the zero levels
of each survey have not been accurately determined. Following
Reich $\&$ Reich (1988), we write the observed sky brightness
temperature $T$ at any point and at a frequency $\nu$ as the sum
of various contributions:

\begin{equation}
T(\nu) = T_{\rm gal}(\nu) + T_{\rm cmb} + T_{\rm ex}(\nu) + T_{\rm
zero}(\nu) \ .
\end{equation}

\noindent In Eq.~(1), $T_{\rm gal}(\nu)$ is the Galactic
brightness temperature; $T_{\rm cmb}$ is the brightness
temperature of the Cosmic Microwave Background ($T_{\rm cmb} =
2.728\pm 0.004\,$K, Fixsen et al. 1996); $T_{\rm ex}(\nu)$ is the
contribution of the unresolved extragalactic sources; $T_{\rm
zero}(\nu)$ is the zero level correction. The contribution of the
extragalactic background is of order of few mK at these
frequencies, i.e. much less than any baseline uncertainty, so that
it can be neglected in the calculation.

\noindent While the zero level of the 1.4 GHz survey has been
accurately set, the zero levels of the 2.7 and 5 GHz surveys have
to be estimated. For this purpose, we make use of auxiliary data
at 408 MHz (Haslam et al. 1982) and 2.3 GHz (Jonas et al. 1998).
The Haslam full-sky survey has an angular resolution of
0.85${\degr}$ and its temperature scale is believed to be accurate
to within 10$\%$. In addition, it is characterized by a well-known
offset of 3 K. We have used the version of the 408-MHz survey
obtained by D.P. Finkbeiner, M. Davis $\&$ D. Schlegel (private
communication). They have removed point sources and destriped the
map by applying a Fourier filtering technique. The 2.3 GHz survey,
which has been carried out with the HartRAO 26-m telescope, covers
67$\%$ of the sky with a resolution of 20$'$. The estimated
uncertainty in the temperature scale is less than 5$\%$ and the quoted 
error in the absolute zero is 80 mK in any direction. The
version of the 2.3 GHz map that we have used is that supplied by
Platania et al. (2003) which has point sources removed and has
been destriped. It is important to point out that the Jonas et al. 
survey is polarization dependent and its absolute zero level 
has been set by comparison with the Haslam et al. 408 MHz survey. 
This fact implies that our determination of the 2.7 and 5 GHz 
zero levels are based both, in practice, on the 408 MHz baseline 
accuracy. At
each frequency (2.7 and 5 GHz), the strategy adopted for setting
the zero level takes into account the specific features of the
survey. The Effelsberg 11-com survey extends up to $b=\pm
5{\degr}$. For these latitutes, in the region $20{\degr} < l <
30{\degr}$, we estimate the contribution from the free-free
emission. This is obtained using the Dickinson et al. (2003)
H$\alpha$ map. Following the lines described in the paper, the
observed  H$\alpha$ is first corrected for absorption by means of
the 100-$\mu$m dust template given by Schlegel, Finkbeiner $\&$
Davis (1998) and the assumption of a non-uniform mixing of ionized
gas and dust {\footnote{In their paper, Dickinson et al. define
$f_{d}$ as the effective dust fraction in the line of sight
actually absorbing the H$\alpha$ and set $f_{d} = 0.33$.}}. The
corrected H$\alpha$ is then converted into free-free emission
using Eq.~(11) in the Dickinson et al. paper. The derived
free-free emission contribution is shown in Fig.~2.

\begin{figure}
\epsfig{figure=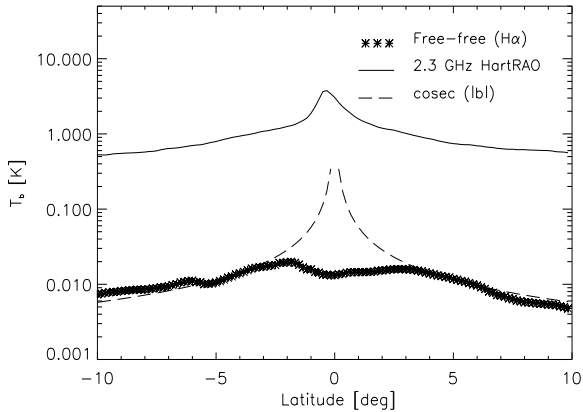, height=6cm,
width=8cm, angle=0} \caption{Free-free emission contribution
(crosses) at 2.3 GHz (observed emission denoted with a solid line)
estimated from the Dickinson et al. (2003) H$\alpha$ map. The
profiles have been obtained by averaging over 10$'$ slices in
latitude from $l = 20{\degr}$ to 30${\degr}$. Shown in the figure
(dashed line), for comparison, is also the best-fitting cosecant
law for the free-free emission distribution. Note the logarithmic
scale on the vertical axis.}
\end{figure}

\noindent As stated in Sect.~2, the H$\alpha$ is not a reliable
indicator of free-free emission for regions close to the Galactic 
plane. However, as the figure shows, at 
$b$=$\pm$ 5${\degr}$ 
the derived free-free emission still agrees with a cosecant law 
while for $|b| < 5{\degr}$ and, in particular, $|b| < 3$--$4{\degr}$, 
a clear departure sets in. Up to $b=\pm 5{\degr}$, the free-free 
emission contribution at 2.3 GHz is negligible with respect to 
the total emission which, consequently, is dominated by 
synchrotron radiation. We make use of the Giardino et al. (2002) spectral
index map to extrapolate the 2.3 GHz data to 2.7 GHz. Details on
the Giardino et al. map will be given in Sect.~4.1. Here we only
emphasize that, although this is generally known to be a pure
synchrotron spectral index map, no actual decomposition of the
thermal and non-thermal emission has been performed in the
original data. However, we have shown that up to $\sim 2\,$GHz,
the observed emission is basically synchrotron radiation for $|b|
>3-4{\degr}$. After convolving the 2.7 GHz data at the 2.3 GHz resolution, a
latitude profile extending out to $b=\pm 5{\degr}$ is obtained by
averaging the total brightness temperature, corrected for the CMB,
over the longitude range 20${\degr}$ to 30${\degr}$. The zero
level at 2.7 GHz is estimated at $b=\pm 5{\degr}$ by extrapolating 
the 2.3 GHz data in frequency (see Fig.~3, top panel). The best estimate of T$_{\rm
zero}(2.7\,\hbox{GHz}) = 0.15 \pm 0.06\,$K which is slightly
larger than the zero level uncertainty of 0.1 K given by Reich et
al. (1990b).

The estimate of the zero level for the Parkes 6-cm survey is more
complicated. The observed region does not extend beyond
$|b|$=1.5${\degr}$. At these low latitudes, the H$\alpha$ cannot
be used to trace the free-free emission. Moreover, the thermal
fraction is expected to represent a significant part of the
emission. We then proceed along these lines. We first notice that,
as reported by Broadbent et al. (1989), the digitized version of
the map has a constant offset added of 1 K. We subtract this
offset from the data. We then consider the region $l=5{\degr}$ to
10${\degr}$, $|b|\le 1.5{\degr}$ in which the number of bad
(=empty) pixels is negligible, even at the scan borders. 
As reported by Haynes et al. (1978), out of 150 blocks in which 
the whole surveyed area has been divided, only 3 show baseline 
discrepancies and, in this case, a baseline adjustment 
to bring these blocks into line has been performed. This means 
that we can evaluate the absolute zero level in the region 
$5{\degr} < l < 10{\degr}$ and confidently adopt such an estimate 
in the region $20{\degr} < l < 30{\degr}$. For $l=5{\degr}$ to 
10${\degr}$, $|b|\le$ 1.5${\degr}$, we fit the spectral indices 
between 408 MHz and 2.3 GHz. We use these spectral indices to extrapolate the 2.3
GHz data to 5 GHz. This operation may formally introduce an error.
However, given the fact that we extrapolate out of the fitting
interval but to a nearby frequency, we can assume such an
error to be small. The 5 GHz data, convolved to the larger 
2.3 GHz beam, are used to derive a latitude profile 
extending up to
$b=\pm 1.5{\degr}$. The zero level is obtained by comparison, at
$|b| = 1.5{\degr}$, with the expected profile from the
extrapolated 2.3 GHz data (Fig.~3, bottom panel). The best
estimate, $T_{\rm zero}(5\,\hbox{GHz}) = 0.10\pm 0.05\,$K, is
comparable to the uncertainty of 0.2 K quoted by Haynes et al. (1978).

\begin{figure}
{\epsfig{figure=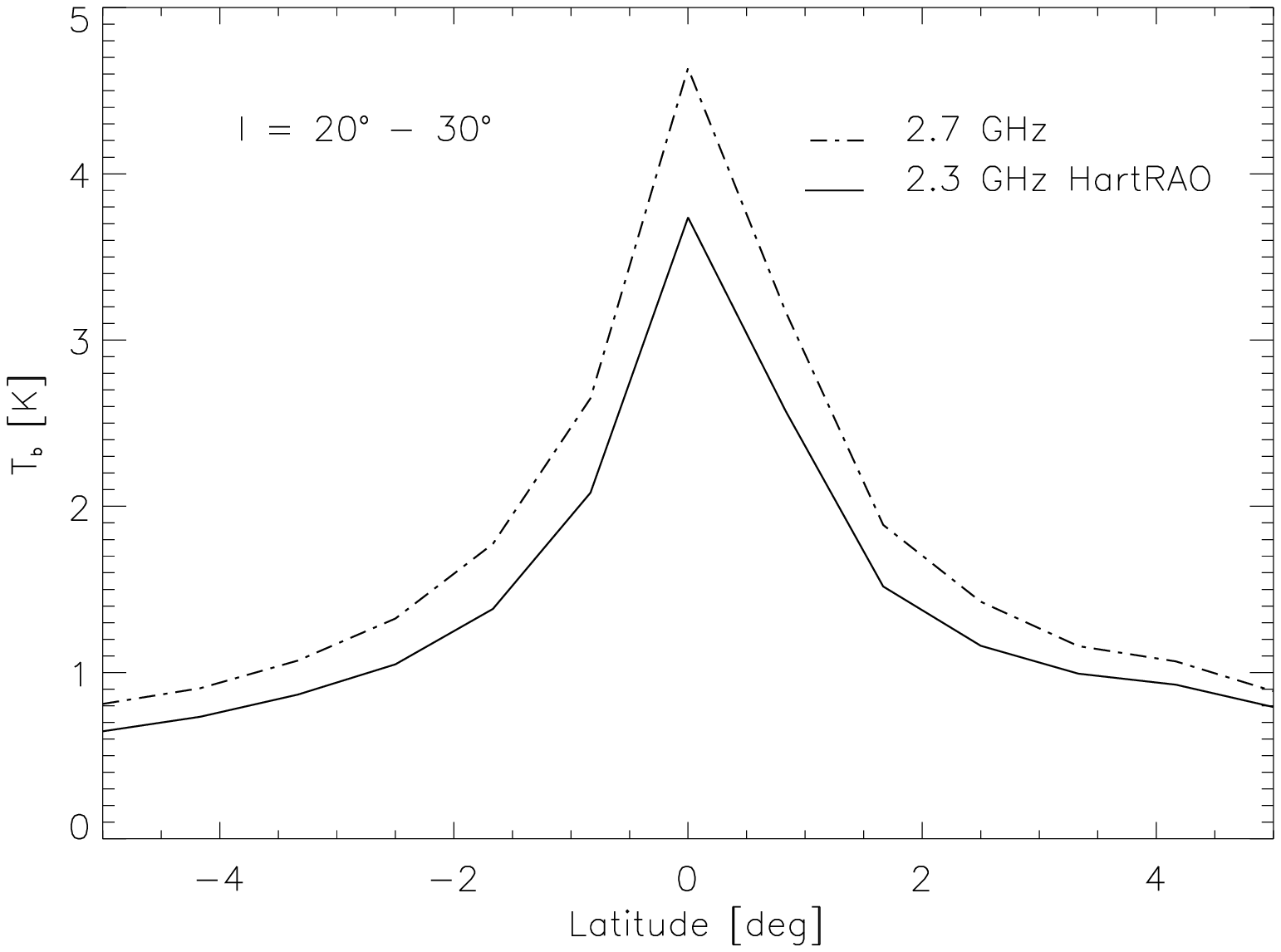,
height=6cm, width=8cm, angle=0}\\
\epsfig{figure=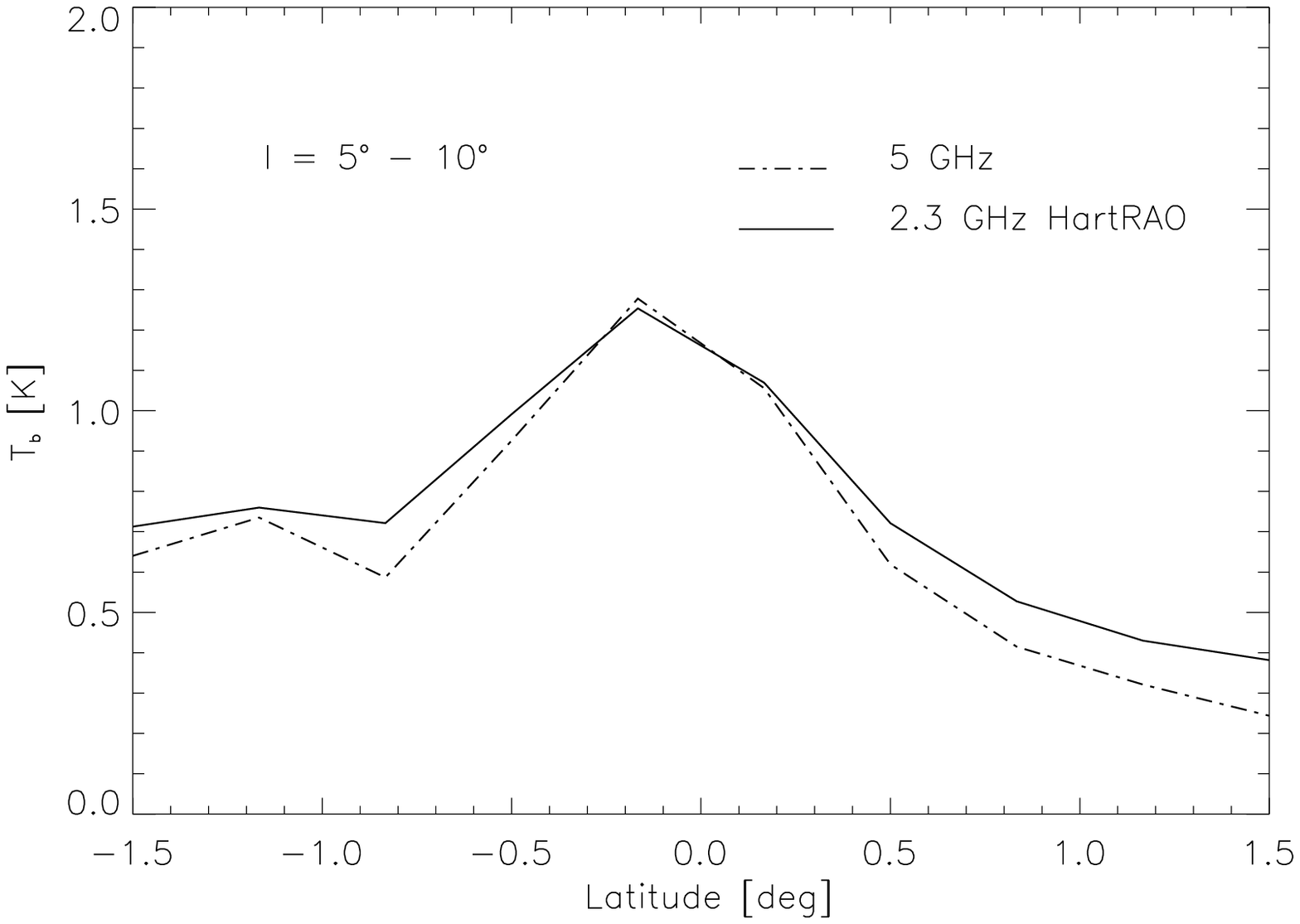,
height=6cm, width=8cm, angle=0}} \caption{The dashed lines show
the latitude profiles for the 2.7 GHz survey (top panel) and for
the 5 GHz survey (bottom panel). The solid lines denote the 2.3
GHz data scaled to the relevant frequency (2.7 or 5 GHz). Profiles
have been obtained by averaging over 20$'$ slices in latitude from
$l = 20{\degr}$ to 30${\degr}$ at 2.7 GHz and from $l = 5{\degr}$
to 10${\degr}$ at 5 GHz.}
\end{figure}

\section[]{Spectral analysis and component separation}

Having set and removed the zero levels at each considered
frequency, we can perform the component separation. At a given
frequency $\nu$, the brightness temperature $T_{\rm gal}(\nu)$ of
a single pixel can be written as:

\begin{equation}
T_{\rm gal}(\nu) = T_{\rm ff}(\nu) + T_{\rm syn}(\nu)
\end{equation}

\noindent where $T_{\rm ff}(\nu)$ and $T_{\rm syn}(\nu)$ are the
free-free and synchrotron emission contributions. By exploiting
the information on the frequency dependence of each component, we
have:

\begin{eqnarray}
T_{\rm gal}(\nu_{1}) & = &
\left(\frac{\nu_{1}}{\nu_{2}}\right)^{\alpha_{\rm gal}} T_{\rm
gal}(\nu_{2})
\\
T_{\rm ff}(\nu_{1}) & = &\left(\frac{\nu_{1}}{\nu_{2}}\right)^{\alpha_{\rm ff}} T_{\rm ff}(\nu_{2}) \\
T_{\rm syn}(\nu_{1}) & = &
\left(\frac{\nu_{1}}{\nu_{2}}\right)^{\alpha_{\rm syn}} T_{\rm
syn}(\nu_{2})
\end{eqnarray}

\noindent In these expressions, $\alpha_{\rm gal}$, $\alpha_{\rm
ff}$ and $\alpha_{\rm syn}$ are, respectively, the Galactic,
free-free and synchrotron spectral indices. The validity of
Eq.~(2) together with Eqs.~(3) to (5) lies in the fact that,
although the sum of two power-laws is not a power-law, it can
always be fitted with a power-law when only two frequencies are
considered. Therefore, by combining Eqs.~(3) through (5), it is
possible to derive the fraction of thermal emission at each
frequency. With some algebra, we obtain:

\begin{equation}
f_{{\rm th}_{\nu_{1}}} =
\frac{1-\left(\frac{\nu_{2}}{\nu_{1}}\right)^{\alpha_{\rm
gal}-\alpha_{\rm
syn}}}{1-\left(\frac{\nu_{2}}{\nu_{1}}\right)^{\alpha_{\rm
ff}-\alpha_{\rm syn}}}
\end{equation}

\noindent We have set $\alpha_{\rm ff}=-2.1$, while for
$\alpha_{\rm syn}$ we built a map of spectral indices as described
in the following section. $\alpha_{\rm gal}$ is computed by
fitting the values of $T_{{\rm gal}(\nu)}$ with $\nu$ respectively
$=1.4$, 2.7 and 5 GHz.

\subsection[]{The synchrotron spectral index map}

One of the major problems in performing a careful separation
between the free-free and the synchrotron emission is the spatial
variation of the synchrotron spectral index. In order to take this
variation into account, we use the spectral index map obtained by
Giardino et al. (2002). This map is constructed by combining data
at 408 MHz, 1.4 and 2.3 GHz and has been convolved to a final
resolution of $10^\circ$ in order to remove the striation due to
the scan-to-scan baseline errors in the input data. The initial
maps have also been median filtered with a a box of kernel of
9$\times$9 pixels (given a pixel size of $\sim$ 13$'$) to suppress
the point source signal.

As discussed in Sect.~3, the Giardino et al. map is not really a
full-sky synchrotron spectral index template. However, for
latitudes above 3--$4^\circ$, it can be regarded as a very good
approximation to that. Below these latitudes, given the difficulty
in obtaining an independent estimate of the free-free emission, it
is not possible to have direct information on the synchrotron. At
the same time, we do not expect a significant gradient in the
synchrotron spectral index distribution within a few degrees from
the plane. For this reason, we built a template of spectral
indices for the selected region of the sky in the following way: 
we assume a spectral index between $b$ = -1.5${\degr}$ to +1.5${\degr}$ 
similar to that between $|b|$ = 4${\degr}$ to 5.5${\degr}$ for our selected 
region $l$ = 20${\degr}$ to 30${\degr}$. The spectral index between  
$|b|$ = 4${\degr}$ to 5.5${\degr}$ is that given by Giardino et al. 
The average spectral index is $\alpha_{\rm syn}=-2.73\pm
0.02$. This is somewhat steeper than the typical spectral index of
supernova remnants (SRNs), $\alpha_{\rm syn} \sim -2.5$ (Green
2004), which contribute significantly to the synchrotron emission
on the plane. However, we note that $\sim 10\%$ of the cataloged
SNRs lie between $|b|$=4${\degr}$ to 10${\degr}$ (Green 2004) so
that their effect is only partly neglected by our analysis. At the
same time, a real steepening of the spectral index is expected
above $\sim$ 1 kpc from the plane (Lisenfeld $\&$ V\"{o}lk 2000).
For a SNR at $D=10$ kpc, we are still within this range at
$|b|=5.5{\degr}$. Given these combined effects, our approach
appears as a viable approximation of the actual distribution of
spectral indices.

Before illustrating the results of the application of Eq.~(6), we
discuss the validity of Eq.~(4) with the assumption $\alpha_{\rm
ff}=-2.1$ over the range 1.4 GHz -- 5 GHz.

\begin{figure}
\epsfig{figure=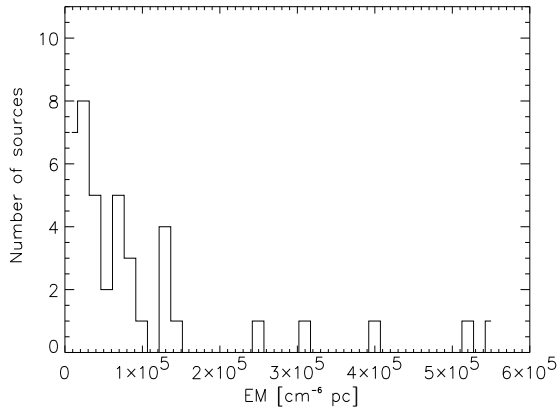, height=6cm, width=8cm,
angle=0} \caption{Emission measure distribution for 41 cataloged
HII regions lying in the range $20{\degr} < l < 30{\degr}$,
$-1.5{\degr} < b < +1.5{\degr}$. Data are from Downes et al.
(1980) and Reifenstein et al. (1970).}
\end{figure}

\subsection{The optical thickness regime for free-free emission in the range 1.4 GHz - 5 GHz}

An important point of the analysis is the assessment of the
emitting conditions of the ionized gas in the frequency range we
are considering and, in particular, at 1.4 GHz. The expression for
the optical thickness $\tau_{\nu}^{\rm ff}$ of thermal
bremsstrahlung in the Altenhoff et al. (1960) approximation is, in
CGS units:

\begin{equation}
\tau_{\nu}^{\rm ff} \simeq 0.08235 T_{e}^{-1.35} \nu_{\rm
GHz}^{-2.1} (\hbox{EM}/\hbox{cm}^{-6}\,\hbox{pc}) \ .
\end{equation}

\noindent If we assume the typical value $T_{e} \simeq 8000$ K
which is common to both the diffuse ionized gas and HII regions,
we see clearly that $\tau_{\nu}^{\rm ff}$ is proportional to the
emission measure (EM). The EM is known to vary widely according to
the form of ionized gas we are considering, i.e. diffuse gas or
discrete, compact sources. For the diffuse gas, we use the
reference value reported by Reynolds (1983) of EM $\simeq$ 9 -- 23
cm$^{-6}\,$pc. For this, at 1.4 GHz we obtain from Eq.~(7)
$\tau_{\nu}^{\rm ff}\sim 1.9\times 10^{-6}$ -- 5.$\times 10^{-6}$
and the optical thin regime applies. Alternatively, if we assume
all the 5 GHz emission on the Galactic ridge to be free-free
emission , then the upper limit to $T_{\rm b, ff}$ is $\le$ 15 K,
i.e. $\tau < 2\times 10^{-3}$ for $T_{e}$ = 8000 K. As for HII regions, of the 102
compact sources lying in our selected coordinate range (see
Sect.~5),  41 have an available estimate of the emission measure.
The distribution of these values is shown in Fig.~4. Most of the
sources are characterized by an emission measure EM $\sim$
$10^{4}$ to $10^{5}\,$cm$^{-6}\,$pc and only 20 percent have higher
values, $> 10^{5}\,$cm$^{-6}\,$pc. In the EM range $10^{5}-10^{6}$
cm$^{-6}$ pc, $\tau_{\nu}^{\rm ff}$ varies in the range $\sim
0.02$ -- 0.2 at 1.4 GHz. Again, since $\tau_{\nu}^{\rm ff} <$ 1, the
assumption $\alpha_{\rm ff}$=-2.1 is correct. Eq.~(7) indicates
that, at 1.4 GHz, $\tau_{\nu}^{\rm ff}$ is bigger than unity for
values of the emission measure $\ge 5\times
10^{6}\,\hbox{cm}^{-6}\,$pc. These values of EM tend to
characterize mainly UCHII whose flux density at these frequencies
is rather faint. In summary, our adopted free-free emission
spectral index appears to be correct over the frequency range 1.4
GHz -- 5 GHz and the error introduced in the analysis by
neglecting the presence of UCHII in our region of the sky is
small.

\subsection[]{The Galactic spectral indices and the component separation}

Before we can make the separation between free-free and
synchrotron emission using the formalism of Eqs.~(3) to (6), we
compute, for each pixel in the selected coordinate range, the
average spectral index, $\alpha_{\rm gal}$, in the frequency range
1.4, 2.7 and 5 GHz. The distribution of $\alpha_{\rm gal}$ between 
1.4 and 5 GHz for the
selected area is shown in Fig.~5. The average value is
$\overline{\alpha_{\rm gal}}=-2.4\pm 0.15$. The latitude variation
in $\alpha_{\rm gal}$ is shown in Fig.~6: there is a marked
flattening of the spectral index on the Galactic plane towards the free-free emission
value of $-2.1$ (with a maximum value of $-2.25$) demonstrating
the dominance of free-free emission there.  At a
Galactic latitude of $b \sim 1{\degr}$ however the synchrotron
fraction is significantly larger with a value of $\alpha_{\rm gal}\sim -2.5$.

\begin{figure}
\epsfig{figure=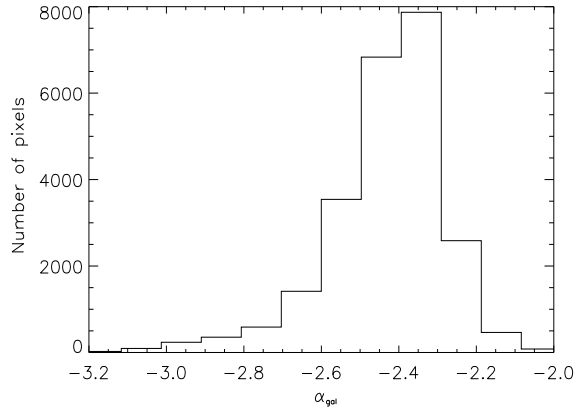, height=6cm,
width=8cm, angle=0} \caption{Galactic spectral index distribution
between 1.4 GHz and 5 GHz. The mean value is
$\overline{\alpha_{\rm gal}} = -2.4\pm 0.15$. }
\end{figure}

\begin{figure}
\epsfig{figure=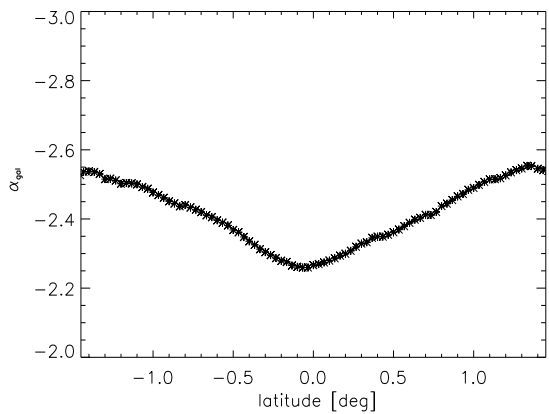, height=6cm, width=8cm,
angle=0} \caption{The spectral index variation with latitude of
the total Galactic emission. The points are averages over
longitudes in the range $20{\degr} < l < 30{\degr}$.}
\end{figure}

By applying Eq.~(6), we can compute the fraction of thermal
emission at 1.4 GHz and 5 GHz. The variation of $f_{{\rm th}_{1.4
{\rm GHz}}}$ and $f_{{\rm th}_{5{\rm GHz}}}$ with latitude is
shown in Fig.~7. The thermal fraction of the total emission clearly 
increases on moving towards the plane reaching a 
maximum value of
82\% at $b=0{\degr}$ at 5 GHz; the corresponding value at 1.4 GHz
is 68\%. Hirabayashi (1974) has carried out 4.2 and 15.5 GHz 
measurements with $10'$ resolution of 9 points on the Galactic equator. Two 
of his points (denoted, in his work, as points A and B) fall in the region analyzed in this 
paper. At  15.5 GHz the free-free emission dominates, the synchrotron 
contribution being at the $\simeq 10\%$ level for $b\simeq 0^\circ$. Our 
approach yields $T_{\rm ff} (15\hbox{GHz}) \simeq 0.103\,$K for point A and 
$T_{\rm ff} (15\hbox{GHz}) \simeq 0.091\,$K for point B, in good agreement (being the difference 
of order of 30$\%$) with 
the measurements by Hirabayashi (1974), corrected for our estimated 
synchrotron contribution.

\begin{figure}
\epsfig{figure=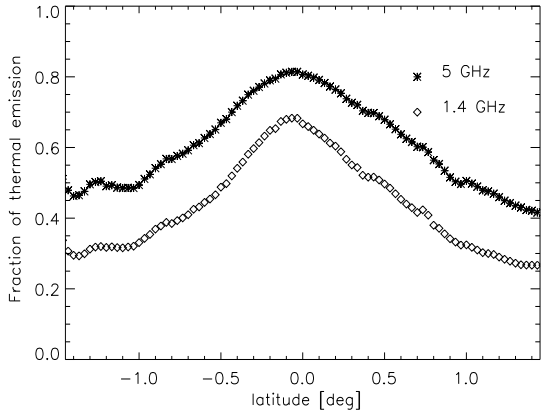,
height=6cm, width=8cm, angle=0}
\caption{Fractions of thermal emission at 5 GHz (upper
curve) and 1.4 GHz (bottom curve).
}
\end{figure}

\begin{figure}
\epsfig{figure=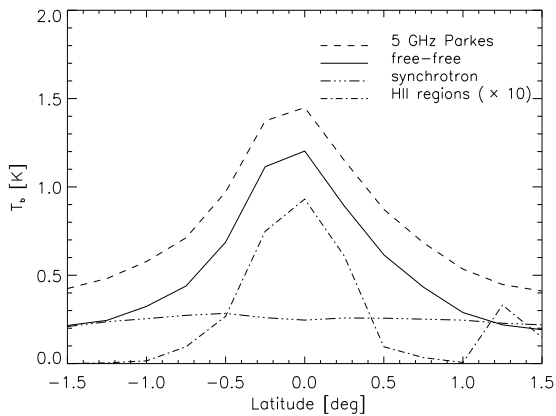, height=6cm,
width=8cm, angle=0} \caption{Latitude dependence of the total
emission (dashed line), and of the free-free (solid line),
synchrotron (double dotted-dashed line) and HII regions
(dotted-dashed line) components for the 5 GHz Parkes survey. The
free-free emission contribution is dominant within $\sim 1^\circ$
from the Galactic plane. At higher latitudes, the synchrotron
fraction increases significantly with respect to the thermal
component. Here, the HII regions profile has been multiplied by a
factor 10. }
\end{figure}

\section{The free-free emission component and the contribution from HII regions}

We now proceed to separate the free-free emission component in the
selected area $20{\degr} < l < 30{\degr}$, $-1.5{\degr} < b <
1.5{\degr}$. Using Eqs.~(3) to (5) pixel by pixel, the fraction 
of thermal emission at 5 GHz can be estimated. The latitude number 
distribution is derived by averaging over longitude strips of the
same size as the pixel side (2$'$). The distributions of free-free
and synchrotron emissions are shown in Fig.~8. A simple gaussian
fit to the free-free emission distribution at $|b| < 1.2^{\circ}$
gives a FWHM of $\sim 1.4^{\circ}$. The actual distribution is
more complex, suggesting a narrow component on the Galactic plane.

A study of HII regions in the area throws light on the situation.
There are 102 compact HII regions in the Paladini et al. (2003)
catalog and 175 ultracompact HII regions in the Giveon et al.
(2004) catalog lying in the selected area. Their latitude
distribution is well-fitted by a gaussian of $\hbox{FWHM} =
0.8{\degr}$. Clearly the free-free emission latitude distribution
could be fitted with a narrow ($\hbox{FWHM} = 0.8{\degr}$)
component comprising $\sim 80\%$ of the peak brightness
temperature. The next question is how much of this emission is
contributed by the $\sim 300$ cataloged HII regions. Fig.~9 shows
the space distribution of the HII regions in the form of a
brightness temperature map at 5 GHz made at 9.4$'$ resolution. The
contribution of the cataloged HII regions to the total free-free
emission distribution, averaged over the range $20{\degr} < l <
30{\degr}$, is plotted in Fig.~8. By summing up the emission over
the pixels for, respectively, the simulated HII-regions map and 
the derived free-free map, we can estimate the contribution of
cataloged discrete sources to the total emission budget. The
result is:

\begin{equation}
\frac{\sum_{j=1}^{N} \hskip 0.1truecm T_{{b}_{{\rm
HII}_{j}}}}{T_{{b}_{\rm ff}}} \simeq 9.2\%\ , \label{eq:HII}
\end{equation}

\noindent where $N=277$, $T_{{b}_{{\rm HII}_{j}}}$ is the
brightness temperature of the $j-th$ HII region and ${T_{{b}_{\rm
ff}}}$ is the brightness temperature of the free-free emission
component at 5 GHz. Of this $\sim 9\%$ contribution to the total
emission, $\sim 8\%$ is due to compact HII regions and only $\sim
1\%$ is given by ultracompact HII regions. It is important to note 
that the Paladini et al. catalog is confusion limited at rather
bright flux densities ($\sim 7\,$Jy at 2.7 GHz), so that the total
contribution of HII regions may be substantially higher.

\begin{figure*}
{\epsfig{figure=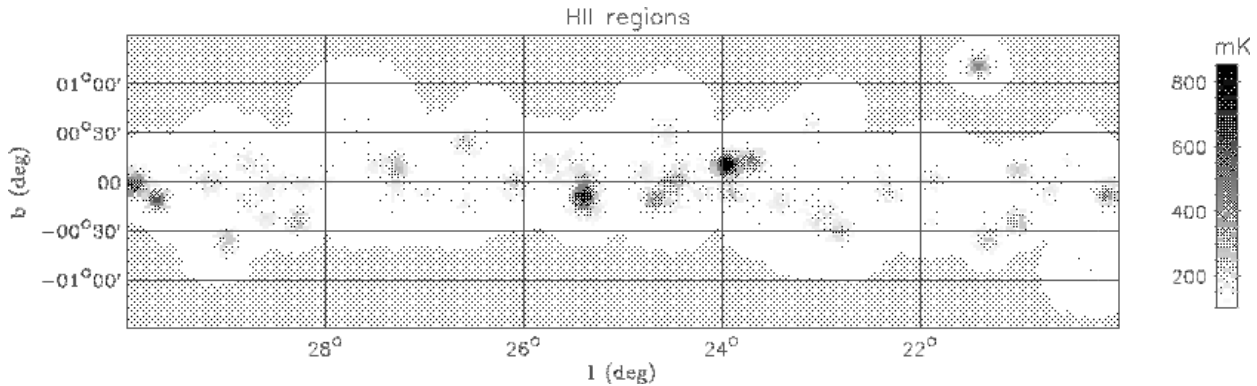,
height=5cm, width=16.5cm, angle=0}} \caption{5 GHz simulated map
for 277 compact and ultracompact HII regions at a resolution of 9.4 arcmin. Flux densities and
angular sizes are taken from Paladini et al. (2003). The colour
scale is logarithmic. }
\end{figure*}

\section{Conclusions}

We have presented the results of an analysis of the diffuse gas
component in the thin layer of warm ionized gas. The analysis,
performed in the coordinate range $20{\degr} < l < 30{\degr}$,
$-1.5{\degr} < b < +1.5{\degr}$, has made use of radio continuum
data at three different frequencies, namely 1.4 GHz, 2.7 GHz and 5
GHz. A decomposition of the total radio emission into free-free
and synchrotron radiation has been carried out by exploiting the
spectral dependence of each component. We present evidence 
that, on a 9.4$'$ angular scale, the free-free emission fraction 
is contributed by both diffuse gas and discrete HII regions. In fact, 
a comparison of the latitude extension of the total
free-free emission distribution with the latitude distribution of
cataloged compact and ultracompact HII regions lying in the
selected coordinate range indicates the presence of a dominant diffuse 
component at 5 GHz. Cataloged HII 
regions contribute only $\sim$ 9$\%$ of
the total emission budget. Although this is only a lower limit
since the catalogs are confusion limited at rather bright fluxes,
it is likely that a large fraction of the the remaining emission
is accounted for by diffuse gas.\\
The errors in deriving the latitude distribution in our 
selected longitude range are mainly due to the uncertainty 
of the synchrotron spectral index which was adopted from the 
Giardino et al. 10 deg resolution spectral index map. Further progress will require 
more frequency data at $\sim$ 10' resolution such as 
will become available from the International Galactic plane 
Survey. A 2-D map of free-free (and synchrotron) emission on the 
Galactic plane should then become a possibility. At the same time, we 
point out that the correlation amoung neighbourig pixels does not 
undermine our results: despite the fact that the oversampling and convolution which have been 
performed on the original data sets introduce a level of correlation of 
order of 1$\%$, given that all the operations involved are linear 
(e.g. averages in longitude), this effect can be neglected. \\
A complementary approach to obtaining the free-free emission 
distribution is to use radio recombination lines (RRLs) which, 
unlike H$\alpha$ emission, are unaffected by dust absorption 
and are not confused by synchrotron emission. The HIPASS 
(Stavely-Smith et al. 1996) and HIJASS (Boyce et al. 2001) surveys are now 
become available. However, an electron temperature is required to convert 
RRL line integrals to free-free brightness; the free-free emission 
distribution derived in the present paper can achieve this.\\
Further ancillary data relating to the electron distribution 
in the Galaxy come from several radio studies, including 
Dispersion Measurements of pulsars and Faraday Rotation measurements 
of pulsars and extragalactic sources. A recent promising initiative 
has been the use of Faraday Rotation of the diffuse Galactic synchrotron 
emission to map the Faraday screen as a function of depth (i.e. frequency) 
(Duncan et al. 1997; Uyaniker et al. 1998, 1999; Gaensler et al. 
2001; Uyaniker et al. 2003). Such an approach gives information about the 
electron and magnetic field distribution.

\section*{Acknowledgments}

The authors thank the referee for helpful comments. They thank 
Giovanna Giardino for providing the processed
version of the 408 MHz and 2.3 GHz maps, as well as for useful
discussions. They also warmly thank Clive Dickinson for
interesting comments and for making available the IDL routine
which allows the conversion of the H$\alpha$ intensity into
free-free emission. Finally, RP thanks Jean-Philippe Bernard for
help in preparing the figures.

\end{document}